\documentclass[11pt]{article}
\usepackage{lmodern}
\usepackage[T1]{fontenc}
\usepackage[utf8]{inputenc}
\usepackage{textcomp}
\usepackage[margin=1in]{geometry}
\usepackage{setspace}
\usepackage{lineno}
\usepackage{microtype}
\usepackage{amsmath,amssymb}
\usepackage{graphicx}
\usepackage{xcolor}
\usepackage{booktabs}
\usepackage{array}
\usepackage{tabularx}
\usepackage{longtable}
\usepackage{authblk}
\usepackage[super,numbers,sort&compress]{natbib}
\usepackage[colorlinks=true,allcolors=blue!55!black]{hyperref}
\usepackage[font=small]{caption}
\usepackage{newunicodechar}
\newunicodechar{σ}{\ensuremath{\sigma}}
\newunicodechar{τ}{\ensuremath{\tau}}
\newunicodechar{β}{\ensuremath{\beta}}
\newunicodechar{α}{\ensuremath{\alpha}}
\newunicodechar{π}{\ensuremath{\pi}}
\newunicodechar{Δ}{\ensuremath{\Delta}}
\newunicodechar{θ}{\ensuremath{\theta}}
\newunicodechar{·}{\textperiodcentered}
\newunicodechar{→}{\ensuremath{\to}}
\newunicodechar{‖}{\ensuremath{\|}}
\newunicodechar{×}{\ensuremath{\times}}
\newunicodechar{≈}{\ensuremath{\approx}}
\newunicodechar{≤}{\ensuremath{\le}}
\newunicodechar{≥}{\ensuremath{\ge}}
\newunicodechar{−}{\ensuremath{-}}
\newunicodechar{●}{\ensuremath{\bullet}}
\newunicodechar{■}{\ensuremath{\blacksquare}}
\newunicodechar{▲}{\ensuremath{\blacktriangle}}
\newunicodechar{½}{\textonehalf}
\graphicspath{{figures/}}
\title{\bfseries Neural noise enables accurate internal simulation of rare events}
\author[1,$\ast$]{Heng Zhang}
\author[2,$\dagger$]{Pawel Herman}
\author[1,$\ast$,$\dagger$]{Zenas C. Chao}
\affil[1]{International Research Center for Neurointelligence (WPI-IRCN), UTIAS, The University of Tokyo, Japan}
\affil[2]{Division of Computational Science and Technology, KTH Royal Institute of Technology, Stockholm, Sweden}
\affil[$\ast$]{Corresponding authors: \texttt{rogerzhangheng@gmail.com}, \texttt{zenas.c.chao@gmail.com}}
\affil[$\dagger$]{Co-senior authors}

\begin{document}
\maketitle

\begin{abstract}
\noindent
The brain needs an accurate internal model of the world to generate
predictions and guide behavior. However, it must estimate the
statistical structure of the environment from limited experience. This
is particularly difficult for rare events, whose observed frequencies in
a limited sample may substantially under- or overestimate their true
frequencies. How the brain constructs an accurate internal model despite
this sampling problem remains unclear. We address this problem using a
Bayesian Confidence Propagation Neural Network (BCPNN) trained on event
sequences from a Markov-chain random walk with controlled event
frequencies. Treating the underlying Markov structure as the ground
truth, we train the network on limited sample of event sequences and
then allow it to generate autonomous replay based on the learned
structure. We evaluate replay fidelity at the levels of both marginal
event frequencies and conditional transition structure. We find that
moderate neural noise, modeled as temporally correlated random
fluctuations in unit activity during replay, is critical for faithful
internal simulation. Without this variability, deterministic replay
systematically under- or overrepresents rare events, whereas moderate
noise restores both their marginal and conditional occurrence. Moderate
noise also broadens the range of parameter values that produce accurate
replay, making the model more robust to parameter variation. Together,
these results support noise-assisted internal simulation as a potential
mechanism for compensating for sampling errors arising from limited
experience. Our model also provides a testable framework for
investigating how altered neural variability may impair internal-model
fidelity in disorders such as Parkinson's disease.
\end{abstract}

\section*{Introduction}

\begin{figure}[htbp]\centering
\includegraphics[width=0.85\linewidth]{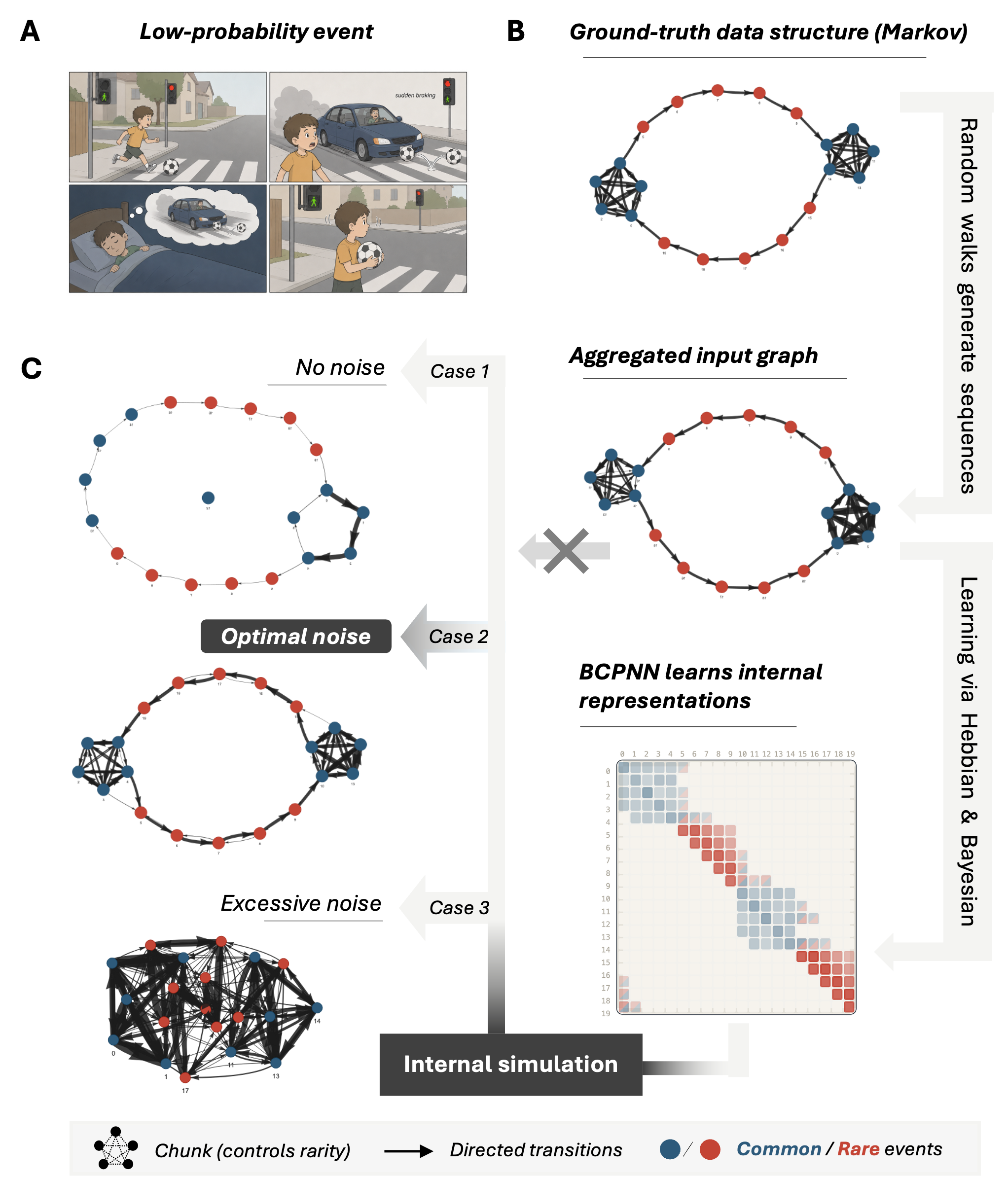}
\caption{\textbf{Overview of internal model simulation of rare events,
and the corrective role of neural noise.} (\textbf{A})~A rare event in
daily life: a child is almost hit by a rushing car, replays the
near-miss offline during sleep, and acts more cautiously at the crossing
afterward, a single experience reshaping prediction and behaviour.
(\textbf{B})~Task and learning: a ground-truth Markov chain of densely
connected common event groups (blue) joined by chained rare event groups
(red) generates training sequences by random walks, and from these
finite samples the BCPNN learns its internal model, the weight matrix
shown and the unit biases, through local Hebbian and Bayesian
plasticity. (\textbf{C})~Internal simulation: cued and left to run, the
trained model replays autonomously under three noise levels. Without
noise (Case~1) the replay mis-expresses the learned structure and
mis-represents the rare events; a moderate, optimal amount of internal
noise (Case~2) recovers the ground-truth structure; excessive noise
(Case~3) washes the structure out.}
\label{fig:motivation}
\end{figure}

Adaptive behavior requires the brain to form an internal model that
accurately captures the statistical structure of its environment.
Predictive-processing theories propose that neural systems learn such
regularities, use them to predict upcoming input, and update internal
representations through prediction errors
\cite{rao1999,friston2010,clark2013,chao2018}. Yet environmental
statistics must be inferred from finite and uneven experience. This
problem is particularly acute for rare events, which may be encountered
only a few times. Their observed occurrence and transition probabilities
can therefore under- or overestimate the underlying environmental
probabilities. Although the present model considers only their
statistical occurrence, Fig. \ref{fig:motivation}A illustrates the broader motivation: a
rare event, such as a near miss, must remain available to internal
simulation if it is to inform future predictions and behavior
\cite{ranganath2003}. An
internal model can consequently fail at two distinct stages: learning
may encode an inaccurate estimate of the environment, and, even when
evidence for rare events is retained in the learned parameters,
internally generated activity may express it inaccurately. Thus,
internal-model accuracy depends not only on what is learned but also on
how the learned structure is sampled during internal simulation. Here,
we focus on this second, circuit-level problem: how autonomous neural
dynamics generate common and rare events with statistics that remain
faithful to the process from which experience was sampled.

Evidence suggests that internally generated neural activity is more than
random background fluctuation. Spontaneous cortical activity can reflect
the statistical structure of stimulus-evoked activity
\cite{berkes2011}, while
hippocampal replay reconstructs rather than simply copies experienced
sequences\cite{gupta2010} and
has been modeled as prioritized sampling that supports memory and
planning\cite{mattar2018}.
Neural-sampling theories similarly propose that stochastic recurrent
dynamics represent probability distributions by visiting alternative
network states\cite{buesing2011}. Such stochasticity is biologically plausible because neural
variability arises at multiple levels, from ion-channel gating and
synaptic release to ongoing network activity
\cite{faisal2008}. Although
noise is often regarded as detrimental, intermediate levels of noise can
facilitate neural computation
\cite{mcdonnell2011}. In
recurrent attractor networks\cite{khona2022}, these observations suggest a specific hypothesis:
deterministic dynamics may collapse onto a restricted set of dominant
trajectories, omitting some low-support alternatives while repeatedly
visiting others. Moderate variability may permit transitions among
learned alternatives, whereas excessive variability may overwhelm the
learned structure. Whether a bounded level of neural noise enables rare
events to be generated with appropriate frequencies during internal
replay remains unclear.

We test this hypothesis using a controlled temporal environment and a
biologically motivated recurrent model (Fig. \ref{fig:motivation}). The environment is
defined by a first-order Markov chain that combines frequently visited
probabilistic communities with rarely visited deterministic chains. Its
stationary event probabilities and transition structure provide an
analytically defined ground truth, while finite random walks over the
chain provide the network's training experience. We use a Bayesian
Confidence Propagation Neural Network (BCPNN), a recurrent attractor
network in which local Hebbian--Bayesian learning stores marginal
activation probabilities and pairwise temporal associations in neuronal
biases and synaptic weights
\cite{sandberg2002,martinez2019} (Fig. \ref{fig:model}). After
training, we hold the learned weights and biases fixed, remove the
external input, and allow the network to generate autonomous replay.
Here, replay refers operationally to autonomous model-generated
sequences, without assuming a particular brain region or behavioral
state. Neural noise is introduced only during replay as independent,
temporally correlated Ornstein-Uhlenbeck fluctuations in the support of
each event-selective unit, the state variable that determines its
activation. We evaluate the resulting replay against ground truth at \textit{two
levels}: the deviation in aggregate rare-event occurrence and the
divergence of the distribution across four transition classes (Fig. \ref{fig:taskmetrics}). This
design isolates the functional expression of learned structure and tests
whether neural variability improves replay fidelity without further
learning or parameter adjustment.

We find that deterministic replay can collapse into periodic attractor
trajectories and either under- or overrepresent rare events, even when
evidence for the corresponding transitions remains in the learned
parameters (Fig. \ref{fig:sim}). Moderate neural noise reduces both forms of error, improving
rare-event occurrence and transition-class fidelity, whereas excessive
noise degrades the learned temporal structure. Moderate noise also
broadens the range of parameter values that support accurate replay
across the tested rarity levels, making fidelity less sensitive to
variation in circuit  (Fig. \ref{fig:result}). These findings support noise-assisted
internal simulation as a generative-sampling layer that complements
experience-dependent learning within predictive processing. Plasticity
stores predictive structure, while appropriately scaled neural
variability allows recurrent dynamics to express that structure without
collapsing onto a restricted set of trajectories. Because the learned
parameters remain fixed during replay, the present results do not show
that noise improves learning itself. Instead, they generate the testable
prediction that, if internally generated trajectories subsequently
engage plasticity, noise-assisted replay could contribute to further
refinement of the internal model. Finally, because Parkinson's disease
is associated with impaired use of probabilistic priors and altered
aperiodic neural activity\cite{perugini2016,gerster2025} (Fig. \ref{fig:disease}), our model provides a testable framework for investigating how
altered neural variability may impair internal-model fidelity in this
disorder.

\section*{Results}

\subsection*{Task design and two-level occurrence fidelity}

A temporal experience can be described as a stream of discrete events,
each occurring with a characteristic frequency, its \emph{occurrence}
\cite{kurby2008,baldassano2017}. For
clarity we take events to be discrete and non-overlapping, one per time
step. Among such events, those of low occurrence, the rare events, are
of particular interest. One might instead single them out by their
importance, as with a seldom-encountered danger, yet importance and
reward already presuppose an estimate of how often an event occurs.
Occurrence is therefore the minimal quantity on which rarity, and later
importance, rests, and it is the quantity we track throughout.

Studying how occurrence is learned in the internal model requires a
generator whose event occurrences can be controlled. Among the standard
descriptions of a temporal stream, including \(n\)-gram statistics,
hidden Markov models, and successor representations, we adopt the most
elementary: a first-order Markov chain, whose event occurrences stay
simple to compute and control exactly. This choice supplies both
quantities of interest from a single object: a stationary distribution
over events (their marginal occurrence) and a transition matrix (their
conditional occurrence).

\begin{figure}[tbp]\centering
\includegraphics[height=0.80\textheight,keepaspectratio]{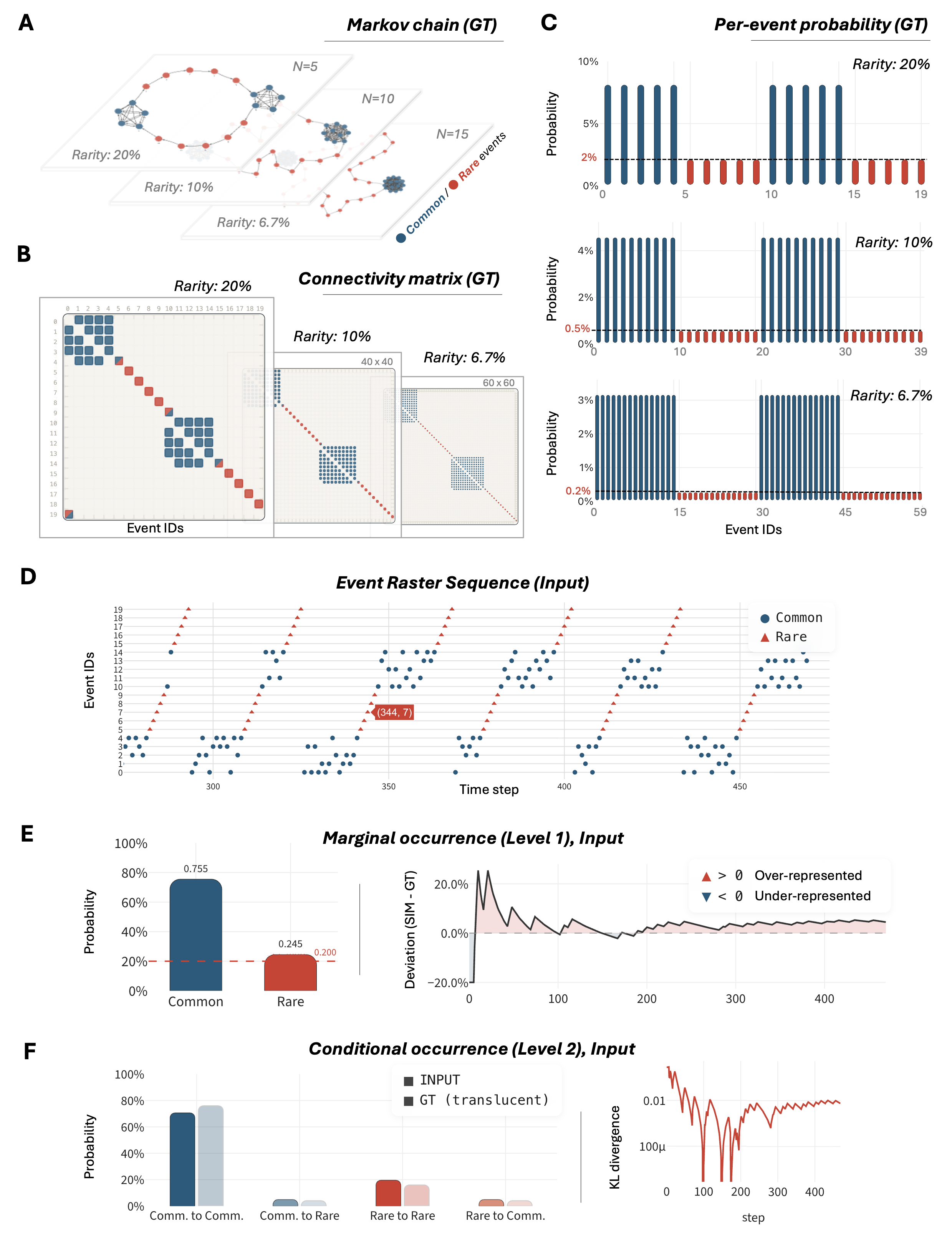}
\caption{\textbf{Task design and two-level occurrence fidelity.}
(\textbf{A}) The ground-truth (GT) Markov chain, shown at three rarity
levels (\(N = 5,10,15\); rare share 20\%, 10\%, 6.7\%); common/rare
events are blue/red. (\textbf{B}) The GT connectivity matrix; dense
diagonal blocks (blue) mark common cliques, and sparse red entries mark
rare chains. (\textbf{C}) Per-event stationary probability, the marginal
occurrence; the dashed threshold refers to GT occurrence of a rare
event. (\textbf{D}) A representative input sequence, shown as an event
raster. (\textbf{E}) Level-1 (marginal) occurrence: the common and rare
occurrence shares are shown, with the GT rare share marked by the dashed
line, and the signed \emph{occurrence deviation} from the GT is
displayed over time, upward red markers denoting over-representation
(\(> 0\)) and downward blue markers under-representation (\(< 0\)).
(\textbf{F}) Level-2 (conditional) occurrence: the four transition-type
occurrences are displayed for the input against the GT (translucent),
and their \emph{occurrence divergence}, a KL divergence, is shown over
time. Definitions are given in \nameref{sec:methods}.}
\label{fig:taskmetrics}
\end{figure}

The transition structure of such a chain can then determine which events
are frequent and which are rare. Two forms of temporal structure have
been characterized in statistical learning in previous studies:
deterministic sequences, in which a fixed order recurs
\cite{saffran1996}; and
probabilistic communities, in which events are grouped by co-occurrence
\cite{schapiro2013,asabuki2020}. Building on
benchmarks that unify these two forms
\cite{vargas2021,zhang2023}, we combine them
in a single cyclic generator
(Fig.~\ref{fig:taskmetrics}A,B), which we take as the
ground truth: densely connected groups of events, visited often,
alternate with sparse deterministic sequences, visited rarely. Rarity is
a structural property of the ground truth rather than an externally
imposed label, as we establish analytically in \nameref{sec:methods}. Although the
grouping is not itself an object of evaluation, we introduce it to
organize the events and to render their occurrence controllable.
Specifically, a single parameter, the size \(N\) of these groups, sets
the degree of rarity: a rare event is visited \(N - 1\) times less often
than a frequent one, so that its occurrence approaches \(1/N\)
(Fig.~\ref{fig:taskmetrics}C). We examine three
settings, \(N = 5,10,15\), corresponding to rare-event shares of 20\%,
10\%, and 6.7\%, spanning a graded range of rarity of the kind surveyed
for machine-learning rare-event prediction
\cite{shyalika2024}. Random
walks over the ground truth produce the finite, imperfect experience on
which the network is trained
(Fig.~\ref{fig:taskmetrics}D). Because the rare-to-rare
transitions are deterministic, any departure from them is signal rather
than sampling noise.

We quantify how faithfully a system estimates the ground truth
occurrence at two levels
(Fig.~\ref{fig:taskmetrics}E,F). At the marginal level
(Level~1), we compare the rare-event occurrence with that of ground
truth, an \emph{occurrence deviation} that is positive when rare events
are over-represented and negative when they are under-represented. At
the conditional level (Level~2), we compare the distribution over the
four transition types (common\(\rightarrow\)common,
common\(\rightarrow\)rare, rare\(\rightarrow\)rare,
rare\(\rightarrow\)common) against ground truth, an \emph{occurrence
divergence} measured by a Kullback--Leibler (KL) divergence. Together
these define the \emph{occurrence fidelity} of a internal simulation
process, which is high only when both the occurrence deviation and the
occurrence divergence are small.

\subsection*{The BCPNN internal model and Hebbian--Bayesian learning}

To reproduce how often events occur in the outside world, an internal
model must store not only which events happen but the order in which
they follow one another; we ground such a model in attractor dynamics.
Recall that each event is by definition a Markov state, and we store
each such state as an attractor, a stable activity pattern that a
recurrent network settles into and holds, with one unit active at a time
\cite{hopfield1982,khona2022}. Holding a state,
however, is not enough: events unfold in time, so the model must also
move between attractors, tracing a sequence rather than resting in one.
Classical attractor networks such as the Hopfield model store static
patterns and provide no intrinsic mechanism for the transitions between
them\cite{hopfield1982}. To address
this, we require an attractor network that captures both the events and
the directed order in which they occur.

\begin{figure}[tbp]\centering
\includegraphics[width=\linewidth]{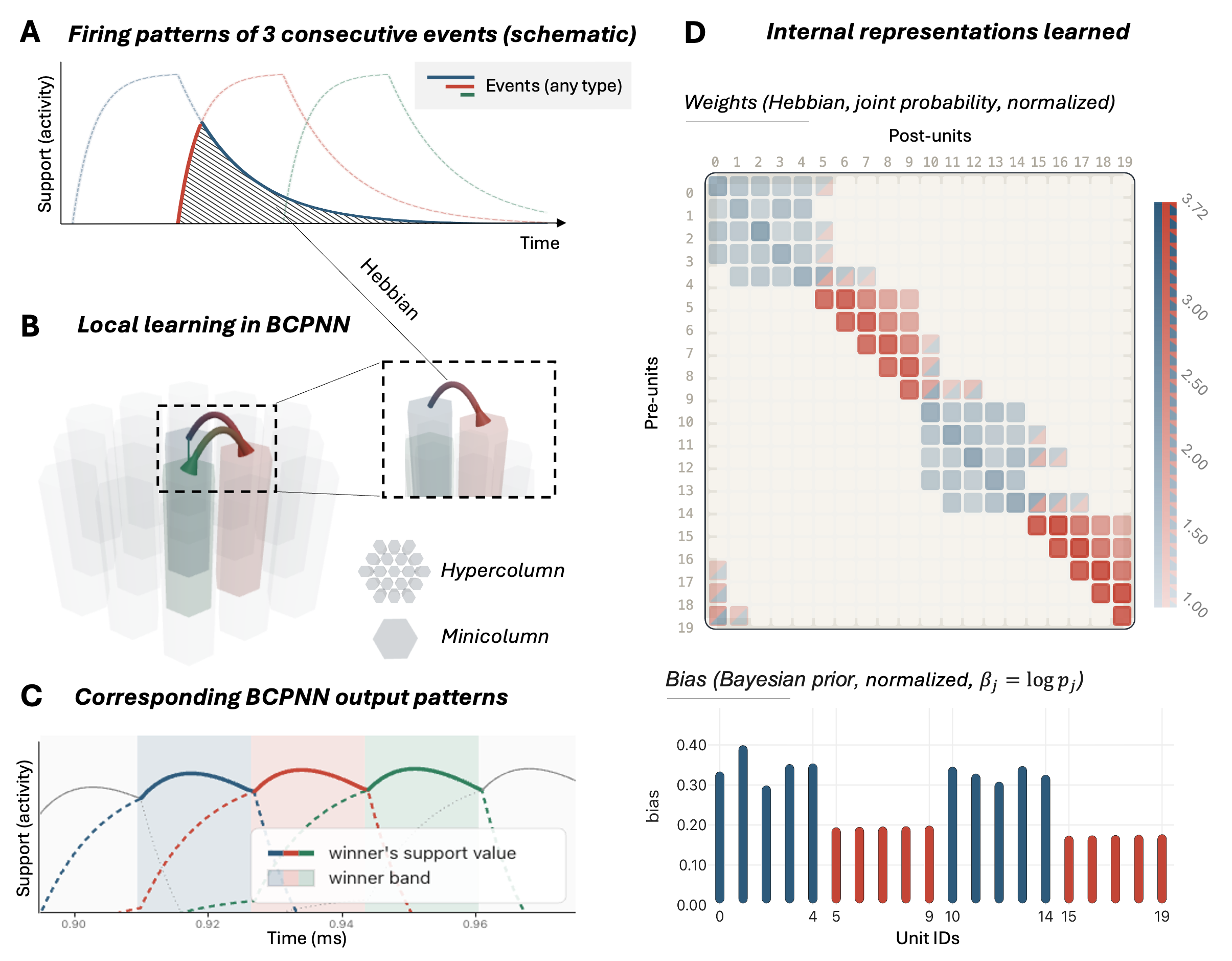}
\caption{\textbf{The BCPNN internal model and Hebbian--Bayesian
learning.} (\textbf{A}) Schematic support (activity) traces for three
consecutive events spanning one event-to-event transition; colour marks
sequence position (any event type), and the shaded region marks the
temporal overlap between consecutive units. (\textbf{B}) Local learning
within a hypercolumn: minicolumns (units) are grouped into one
hypercolumn, and co-activation of consecutive units strengthens the
directed synapse between their minicolumns; the connection shown is
built from the shaded overlap in (\textbf{A}). (\textbf{C}) The
corresponding winner-take-all output, in which each event's support wins
in turn within its ``winner band''. (\textbf{D}) Learned internal
representations after training: the normalized Hebbian weight matrix
(\(w_{\text{ij}}\); pre- versus post-units; block-diagonal frequent
cliques and a directed rare chain) and the normalized Bayesian prior
bias (\(\beta_{j}\), coloured by event type, common in blue and rare in
red). Note that in the weight matrix, cells with \(w < 1\) are hidden
for readability.}
\label{fig:model}
\end{figure}

We use BCPNN, a Bayesian attractor network with local Hebbian learning
\cite{sandberg2002,martinez2022}.
Following the columnar organization of the neocortex, its units are
grouped into a hypercolumn (Fig.~\ref{fig:model}B):
each unit, a minicolumn, represents one event, and its activity encodes
the confidence, a probability, that the event is currently active. A
local winner-take-all (WTA) competition within the hypercolumn
normalizes these activities so that a single unit dominates at each step
(Fig.~\ref{fig:model}C); with one hypercolumn, that
winner is the currently active event. The network dynamics are governed
by a support equation and a competition (see \nameref{sec:methods} for detailed
learning rules):

\begin{align}
ds_{j}& = \frac{1}{\tau_{m}}\left( g_{\beta} \cdot \beta_{j} + g_{w} \cdot \sum_{i}^{}w_{\text{ij}}\, o_{i} - g_{a} \cdot a_{j} - s_{j} + g_{I} \cdot I_{j}(t) \right)\text{\,dt} + \sigma \cdot dW_{j}, \label{eq:support} \\
o_{j}& = \underset{\text{hypercolumn}}{\text{softmax}}(\mathbf{s}). \label{eq:wta}
\end{align}

\noindent Each minicolumn integrates a support \(s_{j}\), the summed synaptic
input to the unit
(Eq.~\ref{eq:support}). The support
combines the prior bias \(\beta_{j}\), the recurrent evidence
\(\sum_{i}^{}w_{\text{ij}}\, o_{i}\) gathered from the active units
through the learned weights, and the external input \(I_{j}(t)\); two
terms return it toward rest, the leak \(- s_{j}\) and the adaptation
current \(a_{j}\), with the membrane constant \(\tau_{m}\) setting the
integration timescale and the gains \(g_{\beta}\), \(g_{w}\), \(g_{a}\),
\(g_{I}\) scaling the prior, recurrent, adaptation, and input terms. We
write the support as an Ornstein--Uhlenbeck process, driven by a
neural-noise term \(\text{σ\,d}W_{j}\) in which \(dW_{j}\) is a per-unit
Wiener increment (variance \(\text{dt}\)) and \(\sigma\) a scalar noise
amplitude (\nameref{sec:methods}). A softmax over the supports within the hypercolumn
then converts them into the activities \(o_{j}\) and enacts the WTA
competition (Eq.~\ref{eq:wta}). Notably, the
adaptation current builds up on the active unit and lowers its support
until control passes to the next, so that the free-running dynamics
trace a sequence of attractors; this adaptation-driven switching
realizes the transitions between events
\cite{martinez2022}.

Learning is local and Hebbian (\nameref{sec:methods}). During training, the external
input clamps each unit to the sampled sequence, injecting the events one
by one (teacher forcing). Consider three consecutive events that
activate three units in turn (Fig.~\ref{fig:model}A;
the traces are schematic, drawn for illustration). A Hebbian rule sets
each weight from the co-activation of the two units it connects: the
more their activity overlaps in time, the stronger the directed
connection (Fig.~\ref{fig:model}B, built from the
shaded overlap in Fig.~\ref{fig:model}A). Because the
events arrive in order, the connection from the first unit to the second
is far stronger than from the first to the third. Meanwhile, the WTA
competition keeps a single unit active at each step, so the output reads
out as the network recalling one event at a time
(Fig.~\ref{fig:model}C).

After a full sequence, the learned weights and biases form the network's
internal representation of the experienced statistics. These two
quantities carry the occurrence structure of the previous section: the
bias \(\beta_{j}\) stores each event's prevalence, its marginal
occurrence, while the weight \(w_{\text{ij}}\) stores the pointwise
mutual information between events, positive when one event follows
another more often than chance and negative below it, its conditional
occurrence. Both identifications follow from writing the support as the
logarithm of a Naive-Bayes posterior, so that the support is accumulated
log-evidence and the activity a confidence, the property that names the
model (\nameref{sec:discussion}). These learned quantities are the network's own
internal model, built from the finite sequence it experienced
(Fig.~\ref{fig:model}D): the weight matrix is
block-diagonal, with strong predictive weights inside the frequent
cliques and a directed chain through the rare sequence, and the biases
are ordered by prevalence, high for common events and low for rare ones.

We train each network by clamping its units to a random-walk sample of
the ground-truth chain, with the neural noise switched off. We then
remove the external input and let the network replay autonomously, now
driven by the learned weights and biases together with the injected
noise \(\sigma\). Once set, every parameter is held fixed across
training and simulation, with two exceptions specified in \nameref{sec:methods} In
what follows we keep the training parameters at their default values
(Table~\ref{tab:params}) and vary only \(\sigma\); the
effect of the training parameters is examined later.

\subsection*{Internal simulation: deterministic replay mis-represents rare events}

We initially evaluate the trained network through live simulation: with
the parameters fixed by training, we cue the network and let it replay
autonomously, monitoring its dynamics over time and tracking the
occurrence of its output against the ground truth as the simulation
unfolds. To make this concrete, we follow two representative instances
at a single rarity (10\%): two networks trained independently on their
own 6000-step random-walk sequences
(Table~\ref{tab:params}), each given the same
common-event cue and then left to run freely.

\begin{figure}[tbp]\centering
\includegraphics[width=\linewidth]{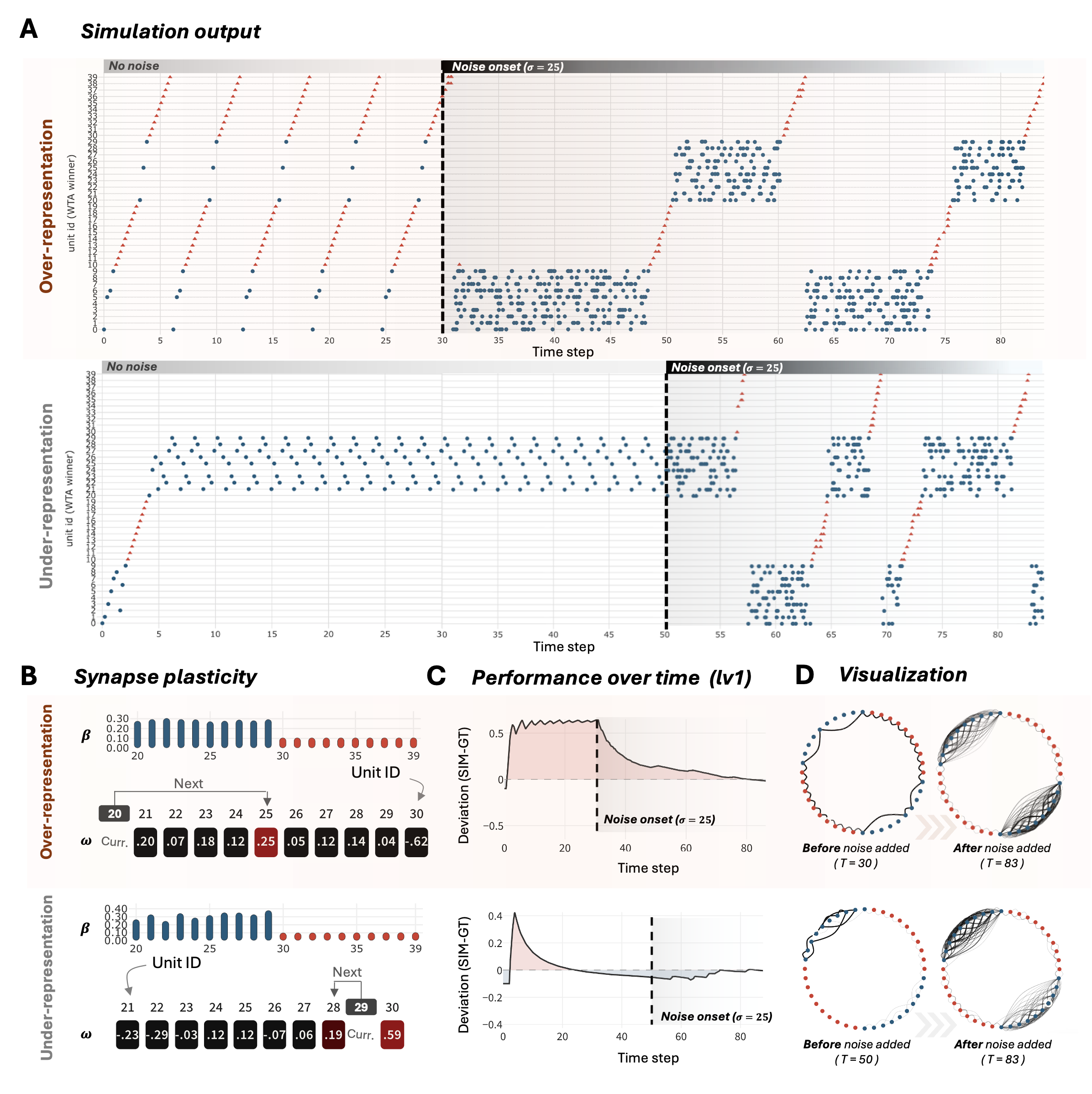}
\caption{\textbf{Deterministic replay and the effect of neural noise.}
Each panel is split into an over-representation case (top, orange tint)
and an under-representation case (bottom). (\textbf{A})
Simulation-output rasters (winning unit id versus time step); in each
case replay is deterministic (\(\sigma = 0\)) until the dashed line,
after which neural noise is switched on (\(\sigma = 25\)). (\textbf{B})
Learned parameters for each case: the bias \(\beta_{j}\) (common blue,
rare red) and the transition weights \(\omega\) from the current unit
(Curr.) to each candidate next unit, on a strength colorbar. In the
over-representation case the largest weight from unit 20 is to unit 25;
in the under-representation case the largest weight from unit 29 is to
the rare unit 30 (\(\omega = 0.59\)), with a smaller weight to the
common unit 28 (\(\omega = 0.19\)). (\textbf{C}) Level-1 (unconditional)
occurrence deviation over simulation time; the dashed line marks noise
onset, set later in the under-representation case so that the
under-representation develops before noise is applied. (\textbf{D})
Transition-count graphs before and after noise onset.}
\label{fig:sim}
\end{figure}

With the neural noise switched off (\(\sigma = 0\)), replay is
deterministic (Fig.~\ref{fig:sim}A): at every step the
winner-take-all selects the unit of highest support, so the trajectory
is fixed by the learned parameters and the cue and settles into a
periodic orbit of attractors, which we call \emph{deterministic replay}.
In neither instance does the output reproduce the ground-truth
occurrence; instead it fails in two complementary ways,
over-representing the rare events in one instance and under-representing
them in the other.

In the first instance, deterministic replay over-represents the rare
events. By taking the single strongest learned transition at each step
and discarding the weaker alternatives
(Fig.~\ref{fig:sim}A,B, top), it skips many common
events and leaves the rare ones over-represented by elimination rather
than amplification, so the Level-1 occurrence deviation rises above zero
(Fig.~\ref{fig:sim}C, top;
Fig.~\ref{fig:sim}D, top).

In the second instance, deterministic replay under-represents the rare
events, trapped in a group of common events that it never leaves for the
rare chain (Fig.~\ref{fig:sim}A, bottom). We identify
one cause in the training experience itself: because the network learns
from a finite random walk, the rare chain is sampled only a few times,
so the rare target is learned with a low prior \(\beta_{j}\). This low
prior then acts through the bias term of the support
(Eq.~\ref{eq:support}). A transition
into the rare chain can carry the strongest learned weight, because the
normalized Hebbian rule
\(w_{\text{ij}} = log(P_{\text{ij}}/(p_{i}p_{j}))\) registers even a
single co-occurrence, yet still be passed over: the support adds the
prior bias \(\beta_{j}\) to this recurrent evidence, and the low prior
of the rare target (small \(\beta_{j}\)) pulls its total support below
that of a common competitor (Fig.~\ref{fig:sim}B,
bottom). This common-to-rare transition is thus vetoed by the low prior,
so the rare chain is never visited and the Level-1 deviation drifts
below zero (Fig.~\ref{fig:sim}C, bottom;
Fig.~\ref{fig:sim}D, bottom).

Thus the two failure modes are two faces of the same support equation:
over-representation is the recurrent term run greedily, while
under-representation is the bias term overriding a strong recurrent
weight. Both produce a large Level-1 occurrence deviation, positive or
negative (Fig.~\ref{fig:sim}C), even though the
internal model encodes the rare structure it fails to express in the
output. Each mode is governed by one of these terms suggests the output
may be sensitive to the corresponding gain, and it raises the question:
how can simulation be corrected so that its output reproduces the
ground-truth occurrence?

\subsection*{Neural noise corrects both failure modes}

Noise is usually treated as a source of error, yet here it is the
corrective ingredient. The support carries a neural-noise term
\(\text{σ\,d}W_{j}\)
(Eq.~\ref{eq:support}) whose amplitude
\(\sigma\) we held at zero throughout the deterministic case; switching
it on perturbs the supports at each step, so the WTA no longer locks
onto the single strongest transition but occasionally takes one of the
alternatives the WTA rule had discarded. This stochastic exploration
loosens the periodic orbit into a trajectory that still follows the
learned supports, and a single such mechanism corrects both failure
modes.

At a moderate noise level (\(\sigma = 25\)), switched on at the dashed
line (Fig.~\ref{fig:sim}), both instances move toward
ground truth. In the over-representation instance, simulation begins to
visit the common events it had been skipping, so the rare events shed
their excess mass and the Level-1 occurrence deviation falls from its
positive plateau toward zero (Fig.~\ref{fig:sim}A,C,
top). In the under-representation instance, replay leaps past the
low-prior barrier into the rare chain it had never entered, so the
deviation rises from its negative plateau toward zero
(Fig.~\ref{fig:sim}A,C, bottom). In both, the
transition-count graphs regain the rare-event edges that deterministic
replay had lost (Fig.~\ref{fig:sim}D). Thus one
undirected source of variability repairs two opposite distortions, and
its magnitude, which we vary systematically next, sets how closely the
output matches the ground-truth occurrence.

\subsection*{Moderate noise recovers rare-event fidelity}

Neural noise corrects both failure modes in the two case studies, but
how much noise is needed, and how reliably the correction holds, remain
open. To quantify this, we sweep the noise amplitude \(\sigma\), the
strength of the neural noise, across its range. So that the result does
not hinge on one training setting, we sweep \(\sigma\) against a
training hyperparameter, \(\tau_{p}\), the timescale over which the
network accumulates the co-occurrence statistics that become its weights
and biases (\nameref{sec:methods}); scoring one autonomous simulation per
\((\sigma,\tau_{p})\) pair yields a two-dimensional grid, here at a
representative rarity (\(N = 5\), \(20\%\) rare).

\begin{figure}[htbp]\centering
\includegraphics[width=\linewidth]{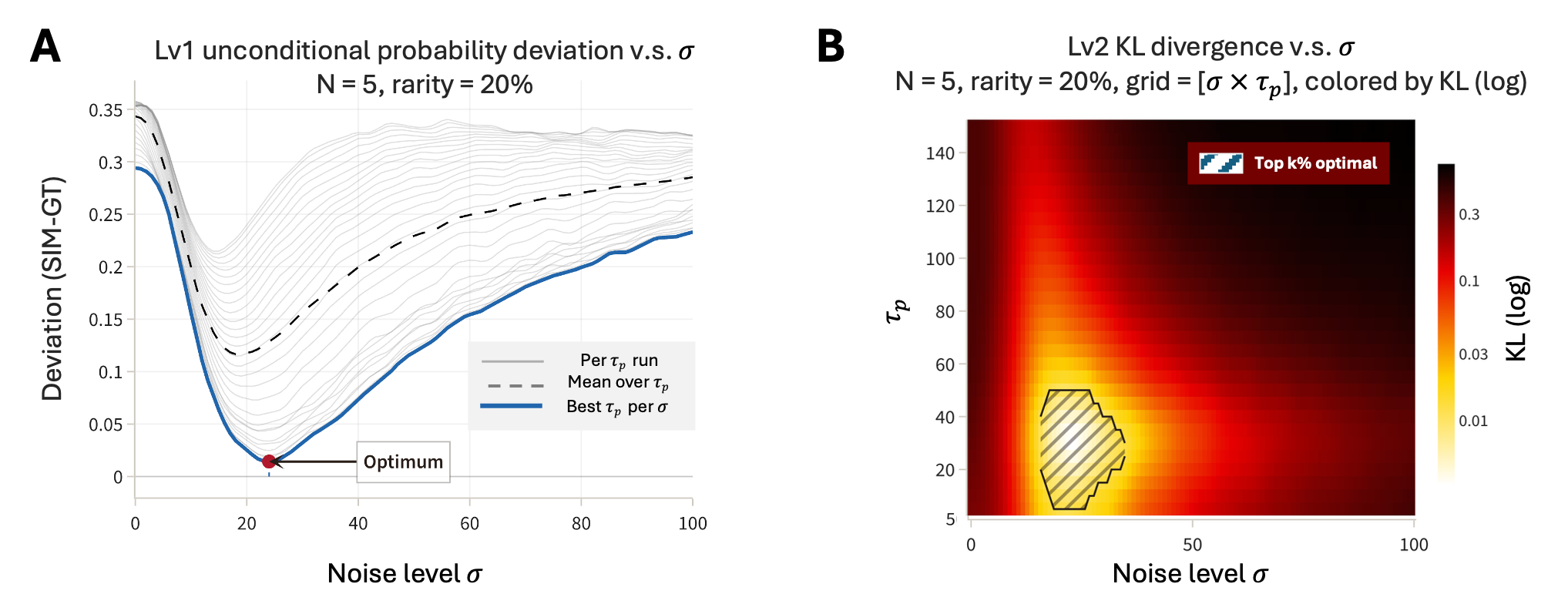}
\caption{\textbf{Moderate replay noise recovers ground-truth structure
at both fidelity levels} (representative case, \(N = 5\), rarity 20\%).
(\textbf{A}) Level-1 unconditional deviation (SIM\(-\)GT) vs noise level
\(\sigma\): thin grey lines are per-\(\tau_{p}\) runs, the dashed line
their mean, and the thick blue line the best \(\tau_{p}\) at each
\(\sigma\); the optimum (red) sits at a moderate, non-zero \(\sigma\).
(\textbf{B}) Level-2 KL divergence over the \(\sigma \times \tau_{p}\)
grid (log colour scale); the hatched region marks the top 5\% most
accurate settings, coinciding with the same moderate-\(\sigma\) band.}
\label{fig:result}
\end{figure}

At Level-1, the unconditional occurrence deviation traces an optimum in
\(\sigma\) (Fig.~\ref{fig:result}A). Deterministic
simulation (\(\sigma = 0\)) leaves a large deviation; as \(\sigma\)
increases the deviation approaches zero, reaching its minimum at a
moderate value (\(\sigma^{*} = 24\), deviation \(0.0143\)); as
\(\sigma\) grows further the deviation worsens again. The individual
\(\tau_{p}\) runs (grey) and their best-at-each-\(\sigma\) envelope
(blue) share this shape, indicating that the optimum is a
moderate-\(\sigma\) band rather than a single-setting artifact. The same
moderate-\(\sigma\) band recovers the conditional occurrence
(Fig.~\ref{fig:result}B): over the
\(\sigma \times \tau_{p}\) grid, the transition-type KL divergence is
minimized in a hatched top 5\% zone that coincides with the Level-1
optimum, indicating that Level-2 tracks Level-1. Therefore, moderate
neural noise recovers the rare-event occurrence at both marginal and
conditional levels, from the same limited experience. Whether this
optimum is specific to \(\tau_{p}\) or generalizes across the model's
other gains and rarities is the tolerance question we take up next.

\subsection*{Tolerance: parameter-agnostic broadening across gains}

The previous section varied the timescale \(\tau_{p}\) at a single
rarity. We further sweep four mechanistically distinct parameters,
\(\tau_{p}\), the prior gain \(g_{\beta}\), the belief-block precision
\(g_{\text{bayesian}}\), and the input gain \(g_{I}\), against
\(\sigma\) across three rarities, scoring both occurrence levels
(Fig.~\ref{fig:grid}).

One of the key findings is that moderate noise broadens the parameter
regime that supports an accurate internal model, such that accurate
representations become less sensitive to parameter selection. In other
words, adding noise improves the \emph{tolerances} of the system.
Specifically, no setting of these parameters is accurate at both
occurrence levels without noise. Once \(\sigma\) is raised to a moderate
range, an accurate regime appears that spans a range of parameter values
rather than a single point, and it persists across all three rarities
(Fig.~\ref{fig:grid}). Moderate noise therefore does
more than locate an optimum: it opens a band of settings that all
support accurate rare-event replay, so accuracy becomes insensitive to
the precise value of the parameter. This robustness is the tolerance the
noise buys. We further verify that this tolerance does not hinge on the
conventional values at which the other parameters are held: halving and
doubling \(g_\beta\) and \(\tau_p\) leaves the correction intact (Appendix, Fig.~\ref{fig:robust}).

Among the four grid searches shown, the prior gain \(g_{\beta}\) behaves
differently from the rest. It sets directly whether rare events are
over- or under-represented(Fig.~\ref{fig:grid}B, top):
the Level-1 deviation is positive at low \(g_{\beta}\) and negative at
high \(g_{\beta}\), with the near-zero band rising as \(\sigma\) grows.
This is consistent with a simple mechanism, in that noise drives the
winner-take-all toward a uniform choice and a larger prior gain restores
each event's marginal share, so the marginal can be held near zero at
any noise level. The transition structure cannot: raising \(g_{\beta}\)
at high noise recovers the marginal but not the conditional occurrence
(Fig.~\ref{fig:grid}B, bottom), indicating that Level-2
is the stricter test and cannot be met by tuning the prior alone.
Besides, the input gain \(g_{I}\) carries a different caveat: its
low-value region reflects a training failure rather than a robustness
effect (Fig.~\ref{fig:grid}D) and is set aside.
Overall, the same tolerance holds across four mechanistically distinct
parameters makes it a property of the noisy dynamics, each with a
distinct provisional neuromodulatory counterpart we take up in the
\nameref{sec:discussion}.

\section*{Discussion}\label{sec:discussion}

\subsection*{Summary of the findings}

We proposed that noise-assisted internal simulation is what lets an
internal model faithfully estimate rare events from limited experience:
without it, the model fails even after learning the structure well.
Specifically, we cast rarity as a structural feature of a cyclic Markov
chain, evaluated simulation at two levels of occurrence (marginal and
conditional), and trained an attractor network (BCPNN) on finite random
walks before letting it replay autonomously under controllable neural
noise. Results showed that, on the same trained model, deterministic
(noiseless) replay over- or under-represents the rare events. However,
without any further learning or tuning, adding moderate noise alone
recovers both their marginal and conditional occurrence, while excessive
noise degrades it. We further show that this correction is
parameter-agnostic, broadening the accurate regime across
mechanistically distinct parameters and rarities. In other words, noise
not only improves accuracy but introduces tolerance, making the system
more robust. This runs against the traditional view of noise as
something to be minimized. However, precisely because neural computation
is noisy, a mechanism that exploits noise rather than suppressing it
invites a biological reading, as we discuss below.

\subsection*{Neural noise}

\textbf{Neurons are inherently noisy.} Stochastic vesicle release,
probabilistic channel gating, fluctuating presynaptic rates, and
network-state fluctuations all inject variability, and on the membrane
they can be combined into an effective noise that a leaky integrator
renders as an Ornstein--Uhlenbeck (OU) process with a single amplitude
\cite{faisal2008}. This is the
noise our model uses: \(\sigma\) scales an OU fluctuation on each unit's
support, its activation, rather than on the synaptic weights, following
the same choice as an earlier BCPNN~sequence network that adds noise to
this variable and degrades gracefully as it grows
\cite{martinez2019}. Ours
differs only in being temporally correlated rather than white.

\textbf{Stochastic resonance (SR).} Read as a dynamical effect, our
central result is an inverted-U: rare-event fidelity is recovered at a
moderate \(\sigma\) and lost again once noise grows too large
(Fig.~\ref{fig:result}). This is the signature of SR,
in which an intermediate level of noise lifts a weak signal across a
threshold and ceases to help once it dominates
\cite{mcdonnell2011}. The
same non-monotonic optimum is well documented in the nervous system, in
single neurons\cite{douglass1993} and in network synchrony\cite{ward2006}, and it takes a striking clinical form in a
predictive-coding account of tinnitus, where the brain injects internal
noise to exploit SR and recover faint afferent signals, at the cost of a
phantom percept\cite{schilling2023}. Its closest echoes are other models of internal simulation that,
despite very different principles, arrive at the same balance: a
variational recurrent network reproduces a data distribution faithfully
only at a moderate level of internal stochasticity, collapsing into
deterministic dynamics when it is too low and into unstructured noise
when it is too high\cite{ahmadi2019pvrnn}, and a reinforcement-learning recurrent network transfers
better when its units are stochastic rather than deterministic
\cite{han2020remaster}. Across
these systems a nonzero amount of noise is the operative ingredient;
where that amount sits, and what happens when it drifts, is where a
biological reading becomes concrete.

\textbf{Neural noise does not stay constant; it increases with age.} Our
account gives noise a useful band, so a system whose noise drifts upward
should slide off the optimum and lose the fidelity that moderate noise
restores. The brain is such a system: the aperiodic (\(1/f\)) slope of
cortical activity flattens in older brains, a signature of increased,
less temporally organized firing, and this flattening statistically
mediates age-related decline in working memory
\cite{voytek2015}. Our model
reads this drift as a departure from the corrective band, a possibility
we take up for Parkinson's disease (PD), where aperiodic broadband power
is likewise elevated\cite{gerster2025}. Besides, excess variability is pathological at the cellular
scale as well: a calcium channelopathy that degrades the precision of
Purkinje pacemaking produces cerebellar ataxia
\cite{walter2006}.
Table~\ref{tab:noise} organizes the sources, function,
and pathology of \(\sigma\) across levels of organization.

\subsection*{Neuromodulation through a shared mechanism}

In biology the parameters that shape our internal simulation are
entangled. Here, we read the parameters as facets of one shared
mechanism; each modulator stays distinct because it acts through its own
receptor types and the signalling they trigger
\cite{drion2015}. We offer the
mapping in this spirit, as a lens on how these systems relate. A few
edges are tight enough to name, and
Table~\ref{tab:neuromod} collects them with a
prediction and the disease in which each is perturbed.

Dopamine (DA) offers the tightest reading, and it touches the prior
along two lines. As a gain, tonic DA raises the intrinsic excitability
and input-output slope of cortical neurons through D1 receptors,
deepening attractor wells and sharpening the weight the network places
on its stored prior; in predictive-coding terms this is DA as the
precision on priors\cite{seamans2004,friston2014}. This
maps onto our prior-belief gain \(g_{\beta}\), whose tonic setting
scales that precision\cite{grace1991}.
As a timescale, DA controls how long the plasticity window stays open,
and a model from the same BCPNN lineage tunes its plasticity time
constant under a dopaminergic relevance signal
\cite{fiebig2014,yagishita2014}. Our
probability-trace timescale \(\tau_{p}\) is the direct analogue of that
constant, giving DA a second, slower point of contact with the prior.
Thus, we hypothesize \(g_{\beta}\) as the sharper leg and \(\tau_{p}\)
as its consolidation-side complement.

Acetylcholine (ACh) offers a second reading, as a switch between
encoding and replay. High cholinergic tone drives encoding from external
input; low tone releases the recurrent, internally-driven dynamics that
our simulation runs on\cite{hasselmo2006}. The same switch reaches three of our parameters at once:
cholinergic tone suppresses recurrent evidence (the \(g_{w}\) in
\(g_{\text{bayesian}}\)), raises the gain on afferent input through
nicotinic receptors (our input gain \(g_{I}\)), and relieves adaptation
(our \(g_{a}\)). One modulator thus organizes a coherent
encode-to-replay transition, which places internal simulation (replay)
in the low-ACh states. During learning, however, cholinergic tone cannot
fall too far. Results in Fig.~\ref{fig:grid}D suggests
that a very low tone would pull the input gain \(g_{I}\) down enough to
corrupt the training phase itself, before replay and its noise come into
question.

Noise gives a third reading. Cortical variability enters our model as a
single effective amplitude, our
\(\sigma\)\cite{faisal2008}, and
arousal plausibly sets its level, with performance best at an
intermediate arousal that echoes our own optimum and noradrenaline (NE)
a candidate driver\cite{mcginley2015}. Read this way, \(\sigma\) is the effective activation
variability the model treats as fundamental, and NE, through arousal,
tunes its level. The locus coeruleus co-releases DA into the
hippocampus, so the DA and NE edges may share a source
\cite{takeuchi2016,kempadoo2016}. We flag
this as the most tentative of the three edges.

\subsection*{Disease models}

\begin{figure}[htbp]\centering
\includegraphics[width=\linewidth,height=0.70\textheight,keepaspectratio]{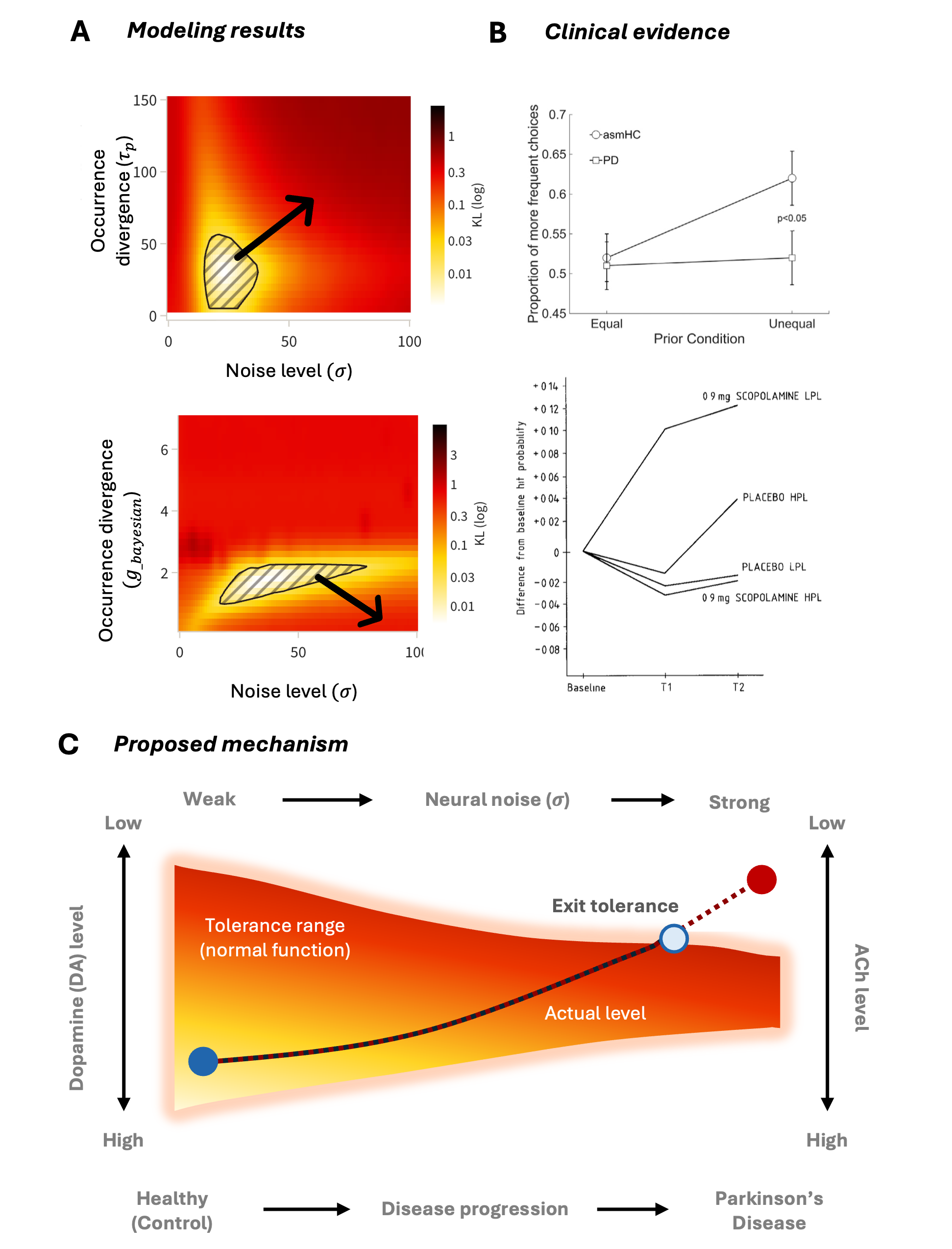}
\caption{\textbf{Parkinson's disease as a convergent test of the
noise-tolerance result.} (\textbf{A})~Model prediction: occurrence
divergence (Level-2 KL) of the internal simulation over the noise
amplitude \(\sigma\) against the dopaminergic parameter \(\tau_{p}\)
(top) and the cholinergic-linked precision \(g_{\text{bayesian}}\)
(bottom); the hatched region is the tolerant zone (top 5\% most
accurate). Black arrows mark the direction in which disease carries the
operating point out of the tolerant zone, toward larger divergence and
greater over-representation of rare events: noise rises (\(\sigma\)
increases, rightward) while the modulatory parameter drifts off its band
(\(\tau_{p}\) upward; \(g_{\text{bayesian}}\) downward, as reduced
acetylcholine lowers \(g_{w}\) and hence \(g_{\text{bayesian}}\)). The
two colour scales differ. (\textbf{B})~Matching human data. Top:
patients with PD fail to shift toward the more frequent (common) option
under sensory uncertainty, unlike age-matched controls (asmHC),
over-weighting the rare; reproduced with permission from Perugini et
al.\cite{perugini2016},
\emph{Current Biology}, Elsevier. Bottom: the muscarinic antagonist
scopolamine (a low-acetylcholine proxy) raises detection at
low-probability locations (LPL) and lowers it at high-probability
locations (HPL), the same over-weighting of the rare; reproduced with
permission from Dunne\cite{dunne1986}, \emph{Psychopharmacology}, Springer Nature, copyright ©~1986
Springer-Verlag. (\textbf{C})~Hypothesis: as the disease advances (left
to right), DA and ACh fall and \(\sigma\) rises together while the
tolerant range of modulation levels narrows; the operating trajectory
follows the healthy state (blue) until it leaves the tolerant range
(exit tolerance) toward the Parkinsonian state (red). All mappings are
provisional, from-modeling hypotheses.}
\label{fig:disease}
\end{figure}

Faithful internal simulation depends on holding the modulatory gains and
the noise level within a tolerant band, so conditions that push them
outside it should degrade the handling of rare events. PD is a natural
first case, since it perturbs more than one of these axes at once. The
loss of DA weakens the prior machinery (the tonic \(g_{\beta}\) and
\(\tau_{p}\)), and patients under sensory uncertainty behave as the
model would predict: they learn the statistics of frequent and rare
events, yet fail to let those priors bias their choices, staying near
chance where controls lean on the common option, which amounts to
over-weighting the rare\cite{perugini2016}. In the same disease, subthalamic recordings show that
aperiodic broadband power, a proxy for asynchronous spiking, scales with
motor severity\cite{gerster2025}, so the noise level \(\sigma\) rises as the prior weakens. Both
moves carry the internal simulation off the tolerant band of our result
(Fig.~\ref{fig:disease}AB, top).

PD also lowers ACh, through degeneration of the basal-forebrain and
pedunculopontine cholinergic neurons
\cite{bohnen2022,bohnen2009}. Read through
the model, a fall in cholinergic tone lowers the recurrent-evidence gain
\(g_{w}\) and, with it, the joint precision \(g_{\text{bayesian}}\); the
operating point then leaves the tolerant island and the network
over-produces rare events. The human counterpart is direct: blocking
cholinergic transmission with scopolamine raises detection at
low-probability locations while lowering it at expected ones
\cite{dunne1986}, the same
over-weighting of the rare that the model shows as
\(g_{\text{bayesian}}\) falls
(Fig.~\ref{fig:disease}AB, bottom).

The two readings converge. DA loss weakens the prior and cholinergic
loss weakens the belief precision, and both, with the rising noise, push
the internal simulation to over-represent rare events, the direction
seen across the two patient studies.
Figure~\ref{fig:disease}C draws this as one trajectory:
as the disease advances, DA and ACh fall and \(\sigma\) rises together
while the tolerant band narrows, until the operating point leaves the
range that supports faithful replay.

The noise axis runs the other way in attention-deficit hyperactivity
disorder (ADHD), where low arousal holds \(\sigma\) below its useful
band and a controlled increase in noise can restore performance
\cite{soderlund2007,helps2014}; the two
conditions sit at opposite ends of that axis. That both extremes fail
underlines the necessity at the heart of the model: faithful internal
simulation depends on a bounded, nonzero amount of noise. Our account
speaks to a single facet, the handling of rare events under uncertainty;
the motor syndrome and the wider pathology of PD remain outside its
scope. We offer the mapping as a testable lens: each edge in
Table~\ref{tab:neuromod} carries a prediction, and PD
is where several of them can be checked at once. Beyond PD and ADHD,
Table~\ref{tab:neuromod} also lists Alzheimer's disease
(AD), where cholinergic loss would move the input and adaptation gains;
we flag it as a from-modeling prediction and leave it to future work.

\subsection*{Biological plausibility of the BCPNN model}

The properties that suit BCPNN to our research problem follow from what
its synapses store. Each unit's activation probability \(p_{i}\) and
each pair's joint probability \(P_{\text{ij}}\) are estimated online
from activity by low-pass probability traces, and the network reads them
out as a bias and a weight. Specifically, the bias
\(\beta_{j} = log(p_{j})\) is the unit's prior, and the weight
\(w_{\text{ij}} = log\left( P_{\text{ij}}/(p_{i}p_{j}) \right)\) is the
pointwise mutual information between units
\cite{martinez2019,martinez2022}. This
pair follows from casting inference as a naive-Bayes classifier, so the
update \(s_{j} = \beta_{j} + \sum_{i}^{}w_{\text{ij}}o_{i}\) is a
log-posterior and learning is local and Hebbian, using only the pre- and
post-synaptic traces\cite{sandberg2002}.

Secondly, unlike the traditional Hebbian rule that ties the weight to
the raw co-occurrence of two units, BCPNN divides that co-occurrence by
the marginals, which is what tunes the rule to rare events. The weight
records how much two units co-occur above chance (\(P_{\text{ij}}\)
against \(p_{i}p_{j}\)), so an association that beats chance registers
even when its absolute frequency is low, and a rare target enters
through a \(- \log p_{j}\) term; the learned \((W,\beta)\) geometry
therefore holds the rare structure of the sequence even where a common
context dominates the traffic. Deterministic replay reads this geometry
off its strongest transitions and loses the rare edges, while a moderate
amount of noise repopulates them, which is the correction our results
turn on. The lineage states the same property directly: even
noise-induced transitions in such a network follow the statistics of the
training set\cite{martinez2022}.

\subsection*{Limitations and future work}

Each simplification in our model points to an extension. First, we
define a rare event by how often it occurs (occurrence). Since frequency
alone can flag a noise outlier as readily as a meaningful event, we plan
to introduce significance to carry importance or reward, weighting what
the network preserves by value. Second, we represent sequences as
first-order Markov transitions, which capture much of the structure we
study yet leave out dependencies a single order cannot express, such as
the oddball regularities of local-global paradigms
\cite{chao2018}. Grouping events
into temporal communities, as in self-organizing chunking
\cite{zhang2023}, would enrich
this structure and let the model act as the cortical component of a
complementary learning system
\cite{mcclelland1995}, where
replay consolidates experience and buffers against catastrophic
forgetting. We also test a single rate-based network in one hypercolumn;
the mechanism already has a spiking realization
\cite{tully2016}, preliminary
runs suggest that adding hypercolumns changes the core result little,
and a broader comparison across models would further locate it. Broadly,
our results identify a bounded, moderate amount of internal noise as an
enabling ingredient of internal simulation: once a network has learned
the statistics of its experience, its own activity fluctuations, held at
the right level, let it replay that experience faithfully with rare
events intact.

\section*{Methods}\label{sec:methods}

\subsection*{The BCPNN model: core}

For a reader new to BCPNN we first state the minimal model; the full
form used in the experiments follows. The network is a single
hypercolumn of \(\text{KN}\) minicolumns, one per event. Each minicolumn
integrates a support \(s_{j}\) that combines the prior bias
\(\beta_{j}\), the recurrent evidence \(\sum_{i}^{}w_{\text{ij}}o_{i}\)
gathered through the learned weights, a leak, and external input, over a
membrane timescale \(\tau_{m}\):

\[\tau_{m}\,\frac{ds_{j}}{\text{dt}} = \beta_{j} + \sum_{i}^{}w_{\text{ij}}\, o_{i} - s_{j} + g_{I}\, I_{j}(t).\]

\noindent The hypercolumn normalizes the supports into activities by a softmax
\(o_{j} = e^{Gs_{j}}/\sum_{i}^{}e^{Gs_{i}}\), whose winner-take-all
(large-\(G\)) limit (Eq.~\ref{eq:wta})
leaves one unit active per step. During training the network estimates
event statistics online: the marginal probability \(p_{j}\) and the
joint probability \(P_{\text{ij}}\) follow low-pass traces on the
probability timescale \(\tau_{p}\),

\[\tau_{p}\,\frac{dp_{j}}{\text{dt}} = o_{j} - p_{j},\text{\quad\quad}\tau_{p}\,\frac{dP_{\text{ij}}}{\text{dt}} = o_{i}o_{j} - P_{\text{ij}},\]

\noindent and the weight and bias are read out as the pointwise mutual information
and the log-prior,

\begin{equation}\label{eq:core_wb}
w_{\text{ij}} = \log\frac{P_{\text{ij}}}{p_{i}p_{j}}, \qquad \beta_{j} = \log p_{j}.
\end{equation}

\noindent Casting inference as a naive-Bayes classifier makes the support a
log-posterior, so learning is local and Hebbian, using only pre- and
post-synaptic traces\cite{sandberg2002,martinez2019}. Because the weight divides co-occurrence by the marginals, an
association that beats chance registers even from a single
co-occurrence, which lets the model store rare structure from limited
experience\cite{martinez2022}.

\subsection*{The BCPNN model: full implementation}

The experiments extend the core with per-term gains, spike-frequency
adaptation, a synaptic z-trace layer, two recurrent channels, and neural
noise. The support becomes the Ornstein--Uhlenbeck form of
Eq.~\ref{eq:support}, which adds the
prior, evidence, adaptation, and input gains
(\(g_{\beta},g_{w},g_{a},g_{I}\)), the adaptation current \(a_{j}\), and
the neural-noise term \(\text{σ\,d}W_{j}\); a single belief-block gain
\(g_{\text{bayesian}}\) scales the prior and evidence together
(Table~\ref{tab:params}) and is held at one except
where swept. The noise makes \(s_{j}\) an Ornstein--Uhlenbeck process,
integrated by Euler--Maruyama with increment
\(\sigma\sqrt{\text{dt}}\,\mathcal{N(}0,1)\) (\(\text{dt} = 1\) ms,
\(\tau_{m} = 10\) ms) and stationary variance
\(\sigma_{\text{out}}^{2} = (\tau_{m}/2)\,\sigma^{2}\). Adaptation is a
low-pass of the unit's own output,

\[\tau_{a}\,\frac{da_{j}}{\text{dt}} = o_{j} - a_{j},\]

\noindent with \(\tau_{a} = 250\) ms in training and \(50\) ms in replay.

The full model inserts a synaptic z-trace between the output and the
probability traces: each output is low-pass filtered into a slow
pre-synaptic and a fast post-synaptic trace, and the probabilities
accumulate these traces. The pre/post asymmetry carries the temporal
order of activation and makes the weights directional
(\(w_{\text{ij}} \neq w_{\text{ji}}\)), the mechanism for
temporal-sequence learning
\cite{martinez2019}:

\[\tau_{z}^{\text{pre}}\,\frac{dz_{i}}{\text{dt}} = o_{i} - z_{i},\text{\quad\quad}\tau_{z}^{\text{post}}\,\frac{dz_{j}}{\text{dt}} = o_{j} - z_{j},\]

\[\tau_{p}\,\frac{dp_{i}}{\text{dt}} = z_{i} - p_{i},\text{\quad\quad}\tau_{p}\,\frac{dp_{j}}{\text{dt}} = z_{j} - p_{j},\text{\quad\quad}\tau_{p}\,\frac{dP_{\text{ij}}}{\text{dt}} = z_{i}z_{j} - P_{\text{ij}},\]

\noindent with the weight and bias read out as in
Eq.~\ref{eq:core_wb}. We run two
pre-synaptic filters, a fast AMPA-like
(\(\tau_{z}^{\text{pre}} \approx 6\) ms) and a slow NMDA-like
(\(\tau_{z}^{\text{pre}} \approx 150\) ms;
\(\tau_{z}^{\text{post}} \approx 5\) ms), giving two weight matrices
whose summed evidence
\(g_{w}\sum_{i}^{}w_{\text{ij}}o_{i} + g_{w,ampa}\sum_{i}^{}w_{\text{ij}}^{\text{ampa}}o_{i}\)
enters the support. This is the z-trace BCPNN of
\cite{martinez2019} (their
Eqs.~6--7); we keep that mechanism and adjust only naming and time
constants. The probability timescale is \(\tau_{p} = 5N - 15\) s (swept
in Fig.~\ref{fig:grid}); logarithms are base~10 and the
probabilities are floored to keep them finite. Parameter values are
collected in Table~\ref{tab:params}.

\subsection*{Task and ground truth}

Experience comes from a directed cyclic graph that interleaves common
and rare structure: \(K = 4\) chunks (two common, two rare) of \(N\)
states each, \(\text{KN}\) states in all, at three rarities
\(N \in \{ 5,10,15\}\). Common (P-type) chunks are densely connected,
each non-exit state projecting to every other state of its chunk, so
they act as probabilistic communities visited often; rare (F-type)
chunks are directed paths, each state projecting only to the next,
traversed deterministically and visited rarely. A single macro-cycle
links the chunks, from a common-chunk exit to a rare-chunk entry to the
next common-chunk entry.

Rarity is a structural property of this graph. Every state has equal in-
and out-degree: a common state has degree \(N - 1\) (reached from, and
projecting to, the other states of its dense chunk), and a rare state
has degree \(1\) (a single deterministic predecessor and successor on
its chain), so the graph is a balanced, strongly connected digraph. For
the random walk that leaves each state along a uniformly chosen
out-edge, \(P(u \rightarrow v) = 1/d(u)\), so every directed edge
carries the same stationary flow
\(\pi(u)\text{\,P}(u \rightarrow v) = c\); the probability entering a
state \(v\) is \(\text{c\,}d_{\text{in}}(v) = \text{c\,d}(v) = \pi(v)\),
and the stationary distribution is proportional to degree,
\(\pi(v) \propto d(v)\). A common state is therefore visited \(N - 1\)
times as often as a rare one. Normalizing over the \((K/2)N\) common
states (degree \(N - 1\)) and \((K/2)N\) rare states (degree \(1\)),

\[\pi_{P} = \frac{2(N - 1)}{KN^{2}},\text{\quad\quad}\pi_{F} = \frac{2}{KN^{2}},\text{\quad\quad}\frac{\pi_{P}}{\pi_{F}} = N - 1,\]

\noindent so the total occurrence of the rare states is exactly
\(\frac{K}{2}\text{N\,}\pi_{F} = 1/N\) (0.20, 0.10, 0.067 for
\(N = 5,10,15\)). Rarity is set by the single parameter \(N\), and this
analytic occurrence is the ground truth against which replay is scored.

We use deterministic chains for the rare chunks, in place of smaller
dense clusters, for three reasons. First, it gives rarity a structural
identity: the rare class is topologically distinct, a sparse chain
against a dense clique, a difference in kind beyond a difference in
size. Second, it gives a closed-form rarity dial,
\(\pi_{\text{rare}} = 2/(KN^{2})\), set by the one knob \(N\). Third, it
fixes a deterministic rare-to-rare ground truth: transitions along the
chain have probability one, so any departure in replay is signal, while
a uniform within-cluster alternative would leave a model error ambiguous
between misfit and sampling noise. We keep \(N\) rare states per chunk
so that the rare-to-rare conditional, the hardest regime to learn,
exists at all. The chain is the simplest structure that makes rarity
both tunable and unambiguously scorable
\cite{schapiro2013,zhang2023}.

Training sequences are single random walks over the graph (6000 steps,
transitions uniform over out-edges), supplying the finite, biased
experience the network learns from; replay is scored against the
analytic ground truth, with no held-out test sequence.

\subsection*{Training}

Training is teacher-forced. Each of the 6000 walk steps is presented as
a one-hot input clamped through the input gain \(g_{I} = 10\), which
drives the winner to the presented event while the probability traces
accumulate; each pattern is held for \(0.1\) s (\(100\) steps of
\(\text{dt} = 1\) ms), about \(6 \times 10^{5}\) steps per network.
Weights and biases are read out from the traces during training and then
held fixed for replay, while the probability traces keep updating. Very
low \(g_{I}\) or very high belief gains corrupt the learned weights
during training itself, a training-time effect (examined in Results)
that we keep out of the replay-tolerance analysis.

\subsection*{Autonomous simulation}

After training we cue the network and let it run without external input.
A random event is clamped for \(0.05\) s (\(50\) steps) to set the
initial state, after which the input is removed and the network evolves
autonomously for \(30\) s (\(3 \times 10^{4}\) steps) under recurrent
evidence, prior, adaptation, and the support noise \(\sigma\). The
winner at each step is read out, and each contiguous dwell contributes
one visit to the occurrence count. For replay the adaptation timescale
is set to \(\tau_{a} = 50\) ms, shorter than the \(250\) ms of training;
this deliberate choice improves how the autonomous run samples
occurrence and treats replay and training as distinct operating regimes.

\subsection*{Evaluation: metrics and grid search}

Replay is scored at two levels of occurrence against the ground truth.
Ground-truth occurrence is the analytic stationary distribution above;
the ground-truth transition structure is the generator's adjacency
normalized to a distribution over its edges. The Level-1
\emph{occurrence deviation} is the signed difference between the
replayed and ground-truth rare-event share,
\(\Delta f = f_{\text{sim}} - f_{\text{gt}}\) with
\(f_{\text{gt}} = 1/N\) (\texttt{simulated\_f\_ratio}); it is zero when
the marginal occurrence matches and its sign shows whether the rare
events are over- or under-represented. The Level-2 \emph{occurrence
divergence} is the Kullback--Leibler divergence from the ground-truth to
the replayed distribution over the four transition classes
(common\(\rightarrow\)common, common\(\rightarrow\)rare,
rare\(\rightarrow\)rare, rare\(\rightarrow\)common),
\(KL(GT\, \parallel \, sim)\) (\texttt{tx4\_kl\_gt\_sim}); collapsing
the transition matrix to four structural classes makes it comparable
across \(N\).

To probe robustness we sweep the noise amplitude \(\sigma\) (0 to 100,
101 values) against each of four training parameters, the
probability-trace timescale \(\tau_{p}\), the prior gain \(g_{\beta}\),
the belief-block gain \(g_{\text{bayesian}}\), and the input gain
\(g_{I}\), at all three rarities, with five independently trained
networks per cell (Fig.~\ref{fig:grid}); the tolerance
width \(\Delta\theta(\sigma)\) is the extent of the parameter interval
whose accuracy stays within \(\varepsilon = 0.05\) of the optimum. The
\(g_{w,overall}\) and \(g_{a}\) axes were swept likewise and appear in
the Supplement. Grid outputs are the \texttt{grid\_search\_*} tables
(inner axis \texttt{sigma\_replay}) in the code deposit; each cell is
five trials seeded as \texttt{global\_seed} \(+\) trial.

\section*{Appendix: Robustness to the choice of fixed parameters}

\begin{figure}[tbp]\centering
\includegraphics[width=\linewidth]{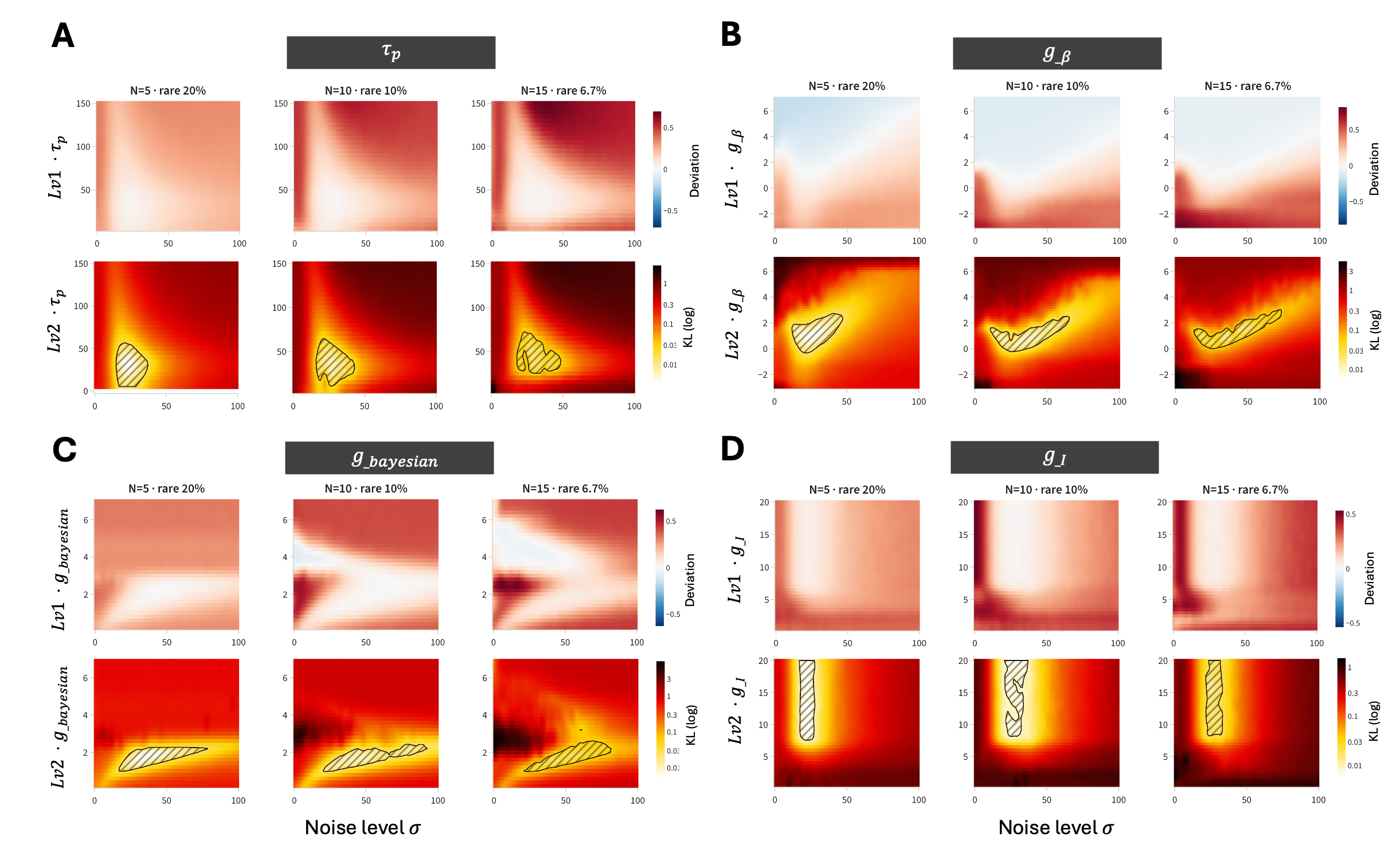}
\caption{\textbf{Moderate noise broadens the tolerant regime across
four mechanistically distinct parameters.} Each block (\textbf{A}
\(\tau_{p}\), \textbf{B} \(g_{\beta}\), \textbf{C}
\(g_{\text{bayesian}}\), \textbf{D} \(g_{I}\)) shows, over the noise
level \(\sigma\) (\(x\)) \(\times\) parameter value (\(y\)) grid, the
Level-1 signed deviation (top row, diverging scale; white \(\approx\)
accurate) and the Level-2 KL divergence (bottom row, log scale; pale
\(=\) accurate), at three rarities (\(N = 5,10,15\); columns). The
hatched region on each Lv2 panel marks the top 5\% most accurate zone
(as in Fig.~\ref{fig:result}). Across all four gains a
moderate-\(\sigma\) band minimizes both metrics and the accurate zone
persists across rarity, suggesting that the noise correction is not tied
to any single dial.}
\label{fig:grid}
\end{figure}

Two of the parameters held fixed in the sweeps of Fig.~\ref{fig:grid} are set by
convention rather than derived: the prior gain \(g_\beta\) = 0.4 and the
probability-trace timescale \(\tau_p\) = 5N − 15. Because each sweep varies
one parameter while holding the others at these values, the tolerance
result could in principle depend on the anchors chosen. To test this, we
repeated three of the sweeps with each fixed value moved to half and
twice its conventional setting, giving 27 independently trained
configurations (three swept parameters × three fixed-value settings ×
three rarities, five trials each), each scored across the full noise
range (σ from 0 to 100 in 101 steps; about 474,000 autonomous replays in
total). Every configuration was seeded so that the training walks depend
on the trial index alone and not on the configuration, so all
fixed-value variants learn from identical random walks and any
difference between configurations is attributable to the fixed value
rather than to sampling. We do not apply this test to the input gain
\(g_I\), whose low-value behaviour reflects a training-time corruption of
the learned weights rather than a replay-tolerance effect (\nameref{sec:methods}); its
panel therefore makes a mechanism claim, not a tolerance claim.

\begin{figure}[htbp]\centering
\includegraphics[width=\linewidth]{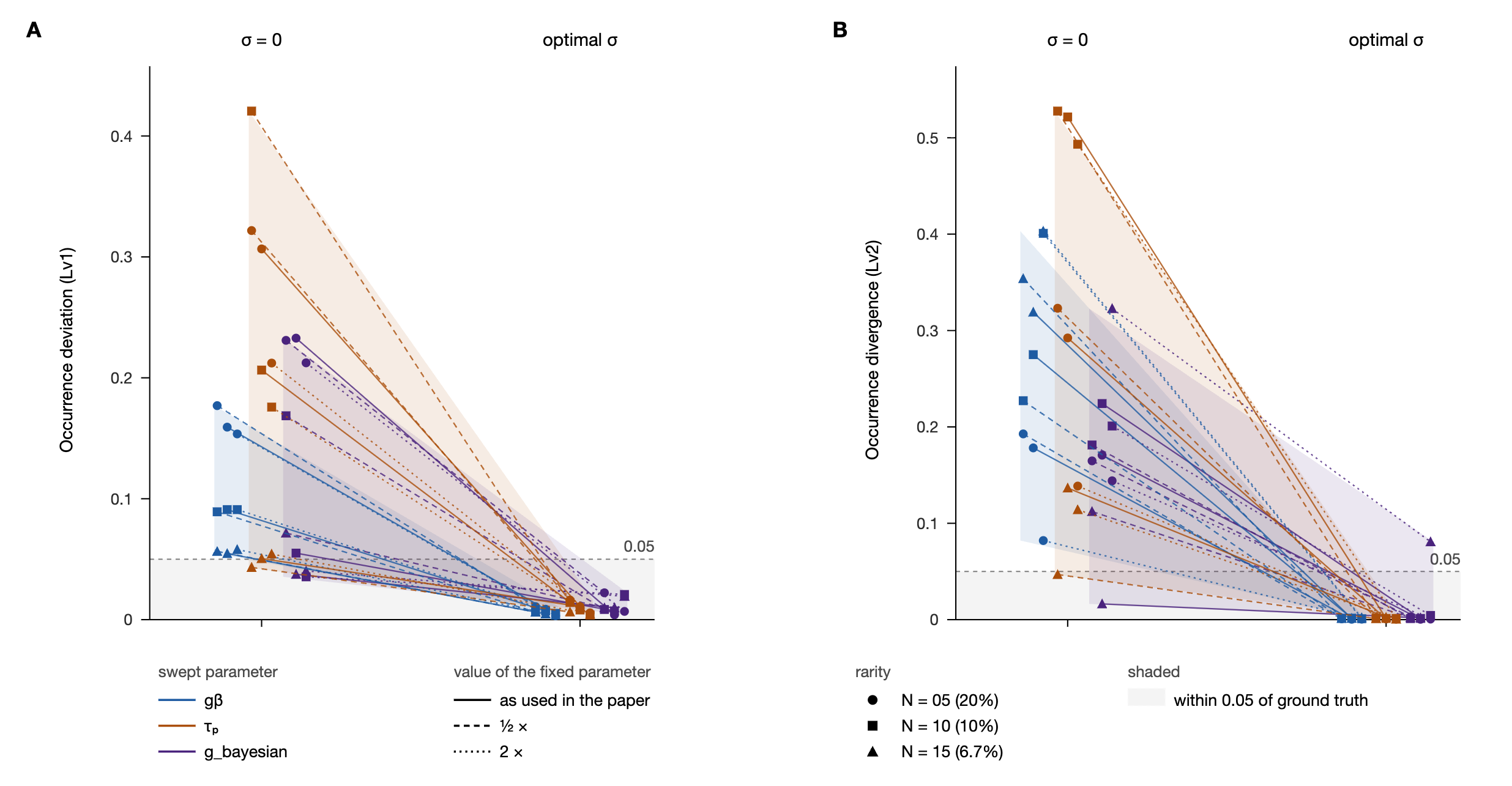}
\caption{\textbf{The correction is not an artefact of the fixed
parameter values.} Each sweep in Fig.~\ref{fig:grid} varies one parameter and holds
the others fixed; two of those values are conventional rather than
derived (\(g_\beta\) = 0.4, \(\tau_p\) = 5N − 15). We repeated three sweeps with each
fixed value halved and doubled, giving 27 independently trained
configurations (3 sweeps × 3 fixed values × 3 rarities, 5 trials each).
(A) Level-1 occurrence deviation and (B) Level-2 occurrence divergence,
each evaluated with no replay noise (σ = 0) and at that configuration's
optimal σ. Without noise the error spans 0.04--0.42 and depends strongly
on the fixed value; at the optimal σ every configuration reaches the
accurate region (A: \textbar Δf\textbar{} ≤ 0.022; B: KL ≤ 0.081). The
shaded band marks 0.05; the shaded wedges span each group's
minimum-to-maximum extent and are not confidence intervals. Horizontal
offsets are for legibility only and carry no quantitative meaning.
Colour indicates which parameter was swept; line style the fixed value
(solid = as used in the paper, dashed = ½×, dotted = 2×); marker the
rarity (● N = 5, ■ N = 10, ▲ N = 15).}
\label{fig:robust}
\end{figure}

The correction is not an artefact of these values. Without replay noise,
the occurrence error spans a wide range across configurations (0.04 to
0.42) and depends strongly on which fixed values were used. Once a
moderate replay noise is present this dependence disappears: at each
configuration's own optimal σ, all 27 configurations converge to the
same accurate region, with a Level-1 occurrence deviation
\textbar Δf\textbar{} ≤ 0.022 and a Level-2 occurrence divergence (KL) ≤
0.081 (Fig.~\ref{fig:robust}). Cherry-picking the anchors would make the result fragile
to these choices; the convergence at the optimum is direct evidence of
the opposite.

The width of the usable noise band is not equally anchor-invariant, and
we report this directly. For the \(g_\beta\) sweep the band is stable (its
extent varies by 2 to 8 percentage points of the swept range across
fixed values); for the \(g_{\text{bayesian}}\) sweep it is moderately sensitive (11
to 20 points); and for the \(\tau_p\) sweep it varies substantially (23 to 50
points), with \(g_\beta\) = 1.2 giving a consistently wider band than the
conventional 0.4 at all three rarities. Performance at the optimum is
unaffected (best \textbar Δf\textbar{} = 0.011 to 0.016 across all \(\tau_p\)
configurations). The defensible statement is therefore specific: the
correction itself is anchor-invariant, whereas the tolerance width for
the \(\tau_p\) sweep is not.

These robustness runs score occurrence with a dwell-aware measure that
resamples the replay trace at a fixed interval before counting, whereas
the \(\tau_p\) and \(g_\beta\) panels of Fig.~\ref{fig:grid} use the earlier visit-count measure.
All 27 configurations here are scored with the same measure and are
internally consistent, so their conventional-anchor lines do not exactly
reproduce the corresponding Fig.~\ref{fig:grid} numbers; the qualitative conclusion
is unchanged.

\bibliographystyle{unsrtnat}
\bibliography{references}

\clearpage
{}

\clearpage
\begin{table}[htbp]\centering
\caption{\textbf{BCPNN parameters and experiment configuration}
(curated; values from
\texttt{main\_bcpnn\_torch\_test\_parallel\_20260605} +
\texttt{config\_network\_params.py}). Symbols defined in \nameref{sec:methods}.}
\label{tab:params}
\small
\begin{tabular}{@{}>{\raggedright\arraybackslash}p{3.2cm} l l >{\raggedright\arraybackslash}p{5.0cm}@{}}
\toprule
Parameter & Symbol & Value & Role / note\\
\midrule

\emph{Task and training} & & &\\
Hypercolumns & --- & 1 & single-hypercolumn network\\
Chunks per cycle & \(K\)\hspace{0pt} & 4 & 2 frequent (P) \(+\) 2 rare
(F), cyclic\\
Chunk size & \(N\)\hspace{0pt} & 5, 10, 15 & rarity \(1/N =\) 20 / 10 /
6.7\%\\
Training sequence length & --- & 6000 & random-walk steps\\
Trials & --- & 5 & independent repeats\\
\emph{Timing and dynamics} & & &\\
Timestep & \(\text{dt}\)\hspace{0pt} & 1 ms &\\
Pattern time & --- & 100 ms & per input pattern\\
Membrane constant & \(\tau_{m}\)\hspace{0pt} & 10 ms & support
integration\\
Adaptation constant & \(\tau_{a}\)\hspace{0pt} & 250 / 50 ms & training
/ replay\\
Adaptation gain & \(g_{a}\)\hspace{0pt} & 10 &\\
\emph{Learning gains} & & &\\
Weight gains (NMDA/AMPA) & \(g_{w},\, g_{w,ampa}\)\hspace{0pt} & 2 / 1 &
recurrent evidence\\
Prior (bias) gain & \(g_{\beta}\)\hspace{0pt} & 0.4 &\\
Input / cue gain & \(g_{I}\)\hspace{0pt} & 10 & cue drive \(+\) training
clamp\\
Belief / evidence scalers & \(g_{\text{bayesian}}\), \(g_{w,overall}\) &
1 / 1 & off at baseline\\
Prob.-trace timescale & \(\tau_{p}\)\hspace{0pt} &
\(5N - 15\)\hspace{0pt} & \(=\) 10 / 35 / 60 for
\(N = 5/10/15\)\\
Competition gain & \(G\)\hspace{0pt} & 1 & strict WTA within
HC\\
\emph{Replay and noise} & & &\\
Replay noise (swept) & \(\sigma\)\hspace{0pt} & 0--100 & 0 \(=\)
deterministic; training \(\sigma = 0\)\\
Recall length & \(T_{\text{recall}}\)\hspace{0pt} & 30 &\\
Cue length / type & \(T_{\text{cue}}\)\hspace{0pt} & 0.05 / random
&\\
\emph{Post-processing} & & &\\
Dwell re-bin & \(dt_{\text{resample}}\)\hspace{0pt} & 0.3 s &
dwell-aware occurrence (splits stuck attractors)\\
Numerical floor & \(\varepsilon\)\hspace{0pt} &
\(10^{- 20}\)\hspace{0pt} &\\
\emph{Grid-search axes} & & &\\
Training gain (outer) & --- & 4 gains &
\(\tau_{p},g_{\beta},g_{\text{bayesian}},g_{I}\) (one per
grid)\\
Replay noise (inner) & \(\sigma\)\hspace{0pt} & 0--100 & 101
points\\
\bottomrule
\end{tabular}
\end{table}

\clearpage
\begin{table}[htbp]\centering
\caption{\textbf{Neural noise across levels of organization.}}
\label{tab:noise}
\small
\begin{tabular}{@{}l >{\raggedright\arraybackslash}p{3.9cm} >{\raggedright\arraybackslash}p{3.9cm} >{\raggedright\arraybackslash}p{3.9cm}@{}}
\toprule
Aspect (\(\sigma\)) & Cellular \& molecular & Circuit \& network &
Systems \& behavioural\\
\midrule

Sources & channel gating, vesicle release & network-state fluctuations &
trial-to-trial variability\\
Function & stochastic resonance in single neurons & network gamma
synchrony & perceptual and clinical SR\\
Pathology & channelopathy: irregular firing & flatter \(1/f\) slope &
disease-level broadband noise\\
\bottomrule
\end{tabular}
\end{table}

\clearpage
\begin{table}[htbp]\centering
\caption{\textbf{BCPNN parameters mapped to neuromodulators and
disease.}}
\label{tab:neuromod}
\footnotesize
\begin{tabular}{@{}l>{\raggedright\arraybackslash}p{2.3cm}l>{\raggedright\arraybackslash}p{3.0cm}>{\raggedright\arraybackslash}p{2.8cm}l@{}}
\toprule
Parameter & Role in BCPNN & Modulator & Hypothesis & Prediction &
Disease\\
\midrule

\(\tau_{p}\)\hspace{0pt} & probability-trace timescale & DA & gates the
plasticity window & slower timescale, slower learning &
PD\\
\(g_{\beta}\)\hspace{0pt} & prior belief (bias) & DA & intrinsic
excitability & less DA, weaker prior & PD\\
\(g_{\text{bayesian}}\)\hspace{0pt} & internal-model precision & DA, ACh
& prior \(\times\) evidence gain & moderate gain best &
PD\\
\(g_{a}\)\hspace{0pt} & spike-frequency adaptation & ACh, NE &
suppresses sAHP and M-current & high \(g_{a}\) frees stuck attractors &
AD\\
\(g_{I}\)\hspace{0pt} & input / cue gain & ACh, NE & gain on afferent
input & low gain corrupts training & ADHD, AD\\
\(\sigma\)\hspace{0pt} & replay-noise amplitude & NE & stochastic
resonance from arousal-linked variability & fidelity peaks at moderate
arousal & PD, ADHD\\
\bottomrule
\end{tabular}
\end{table}

\clearpage

\clearpage

\clearpage

\clearpage

\clearpage

\clearpage

\clearpage

\clearpage

\clearpage
\section*{Data availability}
The data supporting the findings of this study are available from the corresponding authors upon request. Raw simulation outputs, processed data, and analysis scripts used to generate all figures will be made fully available to editors and reviewers upon submission and will be publicly released upon publication.

\section*{Code availability}
All code used in this study is made fully available to editors and reviewers upon submission and will be publicly released upon publication. The computational framework is implemented in Python and distributed as a conda environment (\texttt{bcpnn\_local}). Detailed installation instructions and scripts to reproduce all figures are provided in the repository.

\section*{Ethics statement}
This study involved only computational modeling and simulation. No human participants or animal subjects were involved.

\section*{Acknowledgements}
We thank Dr.~Yohei Yamada and Dr.~Simon Fong for helpful discussions and comments. Computational resources were provided by IRCN. This work was supported by the World Premier International Research Center Initiative (WPI), MEXT, Japan (to Z.C.C. and H.Z.).

\section*{Author contributions}
Z.C.C. and H.Z. conceived the study. H.Z. developed the model, implemented the code, and performed the simulations and analyses. P.H. provided the source code and model details. P.H. and Z.C.C. supervised the work and provided guidance. H.Z. wrote the manuscript with input from P.H. and Z.C.C. All authors reviewed and approved the final version.

\section*{Competing interests}
The authors declare no competing interests.

\section*{Correspondence}
Correspondence should be addressed to H.Z. and Z.C.C.

\end{document}